%% file: main.tex
\documentclass[]{elsarticle}

\usepackage{amsmath}
\usepackage{amsfonts}
\usepackage{amssymb}
\usepackage{fancyhdr}
\usepackage{calc}
\usepackage{url}
\usepackage{tabularx}
\usepackage{graphicx}
\usepackage[usenames,dvipsnames,table]{xcolor}
\usepackage{makeidx}
\usepackage{wrapfig}
\usepackage{subcaption}
\usepackage{hyperref}
\usepackage{url}
\usepackage{verbatim}
\usepackage{natbib}
\usepackage{algorithm}
\usepackage{algorithmic}

\journal{Arxiv}

\newlength{\onecolwidth}
\newlength{\twocolwidth}
\newlength{\colspace}
\begin{document}

\begin{frontmatter}


\title{An adaptive timestepping methodology for particle advance 
in coupled CFD-DEM simulations.}
\author[nreladdesif]{Hariswaran Sitaraman\corref{cor1}}
\cortext[cor1]{Corresponding author}
\ead{hariswaran.sitaraman@nrel.gov}
\author[nreladdesif]{Ray Grout}

\address[nreladdesif]{High Performance Algorithms and Complex Fluids group, Computational Science Center
National Renewable Energy Laboratory, 15013 Denver West Pkwy, Golden, CO, 80401}


\begin{abstract}
An adpative integration technique for time advancement of particle motion in the context
of coupled computational fluid dynamics (CFD) - discrete element method (DEM) simulations is
presented in this work. CFD-DEM models provide
an accurate description of multiphase physical systems where a granular phase exists in an 
underlying continuous medium. The time integration of the granular phase in these 
simulations present unique computational challenges due to large variations in time scales
associated with particle collisions. The algorithm presented in this work uses a local
time stepping approach to resolve collisional time scales for only a subset of particles 
that are in close proximity to potential collision partners, thereby
resulting in substantial reduction of computational cost. This approach is observed 
to be 2-3X faster than  traditional explicit methods for problems that 
involve both dense and dilute regions, while maintaining the same level of accuracy.
\end{abstract}


\end{frontmatter}

\input{introduction}
\input{mathmodel}
\input{results}
\input{conclusions}

\section*{Acknowledgments}
This research was supported by the Department of Energy for project titled ``MFIX-DEM enhancement for industry relevant flows". 
The authors acknowledge helpful discussions with their collaborators at Colorado University, Boulder (Christine Hrenya, Thomas Hauser, Peiyuan Liu, Aaron Holt and Dane Skow), 
Ann Almgren from Lawrence Berkeley National Laboratory, Jordan Musser from National Energy Technology Laboratory and Deepthi Vaidhynathan from 
National Renewable Energy Laboratory.
The U.S. Government retains and the publisher, by accepting the article for publication, acknowledges that the U.S. Government 
retains a nonexclusive, paid-up, irrevocable, worldwide license to publish or reproduce the 
published form of this work, or allow others to do so, for U.S. Government purposes.

\section*{References}

\bibliographystyle{elsarticle-num}
\bibliography{refs}

\end{document}

%% file: introduction.tex
\section{Introduction}

Simulations with coupled computational fluid dynamics (CFD) and discrete element method (DEM) 
are typically used to model physical systems that involve the motion of
particles in an underlying fluid medium. These multiphase systems are observed in several industrial
processes such as fluidized beds \cite{joseph2007,tsuji1993}, riser reactors \cite{lan2009,dutta2012} 
, combustion and reacting systems \cite{kolakaluri2014, capecelatro2015}. The coupled
CFD-DEM approach provides a more accurate representation of these physical systems compared to 
multiphase approaches such as two-fluid models \cite{tsuji1998} where continuum transport equations are solved for the dispersed phase. 
The latter approach requires substantial modeling of closure terms for the interaction between phases \cite{tsuji1998}. 
The CFD-DEM approach, on the other hand, provides an alternate closure for these terms based on 
Newton's laws to govern the motion of particles that represent the dispersed phase. 
The interaction between the particle and continuous phase
requires models only for the fluid induced forces such as lift and drag. In academic settings, 
such simulations are often configured so that the particles represent real particles 
and the interaction with the continuous phase is captured by momentum exchange due to particle drag. 
The continuous phase is treated as usual by discretizing the fluid transport equations for mass, momentum and energy.
However, the more accurate physical representation 
comes with increased computational costs for large scale industrial systems with very small particle sizes. 

One of the computationally expensive facets of a coupled CFD-DEM solver is the time integration of particle
motion. Explicit schemes are typically preferred for time-stepping in DEM due to their accuracy, minimal 
storage and compute requirements \cite{kruggel2008,rougier2004}; implicit schemes in the context of DEM that involve Jacobian computations for 
large number of particle counts (order of $10^6-10^9$) are expensive and infeasible \cite{samiei2012}. 
The use of explicit methods however impose numerical stability and accuracy restrictions on time-step sizes; 
specifically particles that undergo collisions need to be advanced with smaller time-steps compared to non-colliding particles.
The time-step is more often globally set as the limit for accuracy and stability imposed by the collisions and 
is typically orders of magnitude less than that required away from collisions. 
This work addresses this precise issue and provides a strategy to avoid the use of a global conservative
small time-step size for the entire set of particles. A novel time-stepping algorithm 
for CFD-DEM solvers using a partitioning approach that allows for variable time-steps among particles is described and its computational performance is compared against a baseline explicit method, typically used in several
CFD-DEM solvers.

%% file: mathmodel.tex
\section{Mathematical model and numerical methods}

The CFD-DEM solver, MFIX-DEM \cite{syamlal1993}, implemented on adaptive mesh refinement library AMReX \cite{amrex-castro} , will be used in this work. 
MFIX uses an incompressible staggered-grid solver for the continuous phase while particles are transported in a
lagrangian fashion using fluid and body forces (such as drag, pressure gradient and gravity). They also
 undergo collisions, where a soft sphere model such as the Linear-spring-dashpot (LSD) \cite{garg2010} approach is used.
 In most coupled simulations, the fluid time-step size, $\Delta t_f$ 
is larger than the particle time-step, $\Delta t_p$. Therefore, coupled simulations are performed using a subcycling approach as shown 
by the following tasks at each time-step.
\begin{enumerate}
\item{The incompressible Navier-Stokes equations are integrated for a time-step of $\Delta t_f$ for 
the continuous phase.}
\item{Particles are integrated using $\Delta t_p$ time-step size for time $\Delta t_f$.}
\begin{enumerate}
    \item{body forces are calculated using interpolated fluid variables onto particle positions.}
    \item{collisional forces are computed using particle neighbor lists.}
    \item{velocity and position are advanced using an explicit scheme.}
\end{enumerate}
\item{Particle data is deposited onto the fluid grid using deposition kernels and volume averaging.}
\end{enumerate}

The focus of this work is on the time integration aspect of the discrete element method. The particle position and
velocity are typically advanced using an explicit second order velocity Verlet \cite{verlet1967} scheme. In this method, velocity
is advanced at half time levels, $n+1/2$ while position at time level $n+1$ is advanced using this updated velocity. The
numerical scheme can be expressed as
\begin{equation}
\mathbf{v_i}^{n+1/2}=\mathbf{v_i}^{n-1/2} + \Delta t_p \frac{\mathbf{F_i}^n}{m_i};\,\,\,\,\,\,\,\,\,
\mathbf{x_i}^{n+1}=\mathbf{x_i}^n + \Delta t_p \mathbf{v_i}^{n+1/2}.
\label{velverlet-eq}
\end{equation}
Here, $\mathbf{v_i}$, $\mathbf{x_i}$ and $m_i$ represent the velocity, position vector and mass of particle $i$, respectively. 
$\mathbf{F_i}$ represent the total force on particle $i$ computed using position and velocity from the previous time level $n$.

\subsection{Explicit scheme with orthogonal recursive bisection (ORB)}
The time-step size for the explicit method is restricted by the collisional time scale. This time scale depends on the collisional spring constant, 
$k$ and mass of the particle, given by $\tau_p = \sqrt{m_i/k}$. The traditional velocity Verlet 
scheme requires the time-step size, $\Delta t_p$ to be a factor of 5-20X lower than the collision time scale $\tau_p$ for stability
and accuracy \cite{tsuji1993,sullivan2004}. However, this restriction only applies to particles that collide during a given time-step
and need not be applied to dilute zones of the computational domain. An adaptive time-stepping approach where colliding
particles are resolved using collisional time scales while advancing non-colliding particles with larger time-steps can 
provide significant savings on computational costs. This is the motivation for our explicit ORB scheme.

ORB \cite{berger1987,popov2007} is one of the several possible heuristics for partitioning point clouds 
and identifying clusters; k-means \cite{kanungo2002} clustering and Minimum Spanning Tree (MST) \cite{graham1985}
are all possibilities with trade offs between construction/incremental update time vs. effectiveness.
ORB has advantages of being relatively quick and easy to update incrementally and has 
the required heuristic behavior (i.e., it will split the region in half with a cluster on each side) 
when groups of particles are well separated (clustered). When the partitioned domain 
is about the size of the cluster, ORB will split it in half which is 
almost the worst case scenario. The results presented in section \ref{results-sec} 
are sensitive to the interplay between the number of ORB levels and the 
particle cluster size/separation, and we continue to look 
for heuristics that have better balance of the behavior we want with speed of construction. 

\begin{figure}
\centering
\begin{subfigure}{0.4\textwidth}
    \includegraphics[width=0.9\textwidth]{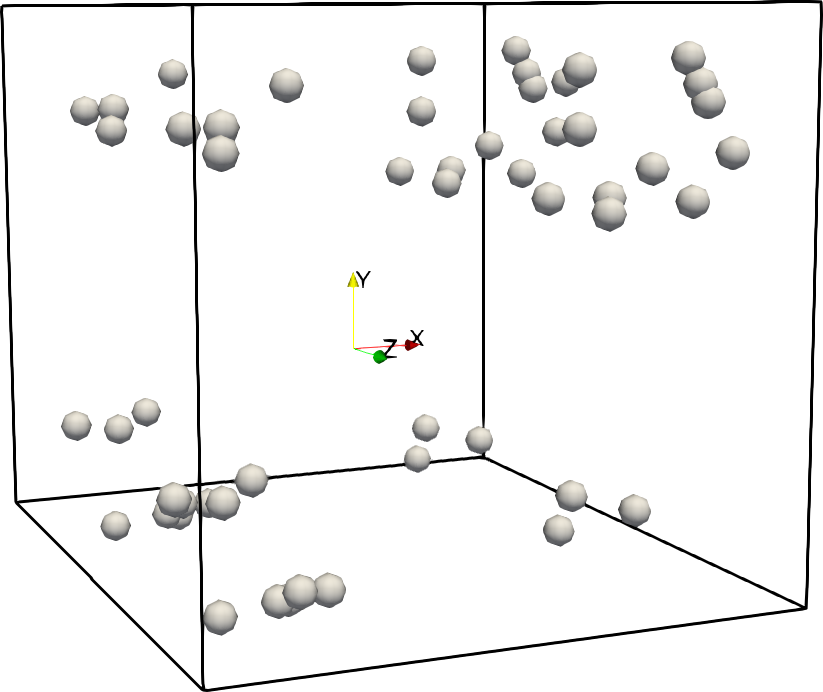}
    \caption{}
\end{subfigure}
\begin{subfigure}{0.4\textwidth}
    \includegraphics[width=0.9\textwidth]{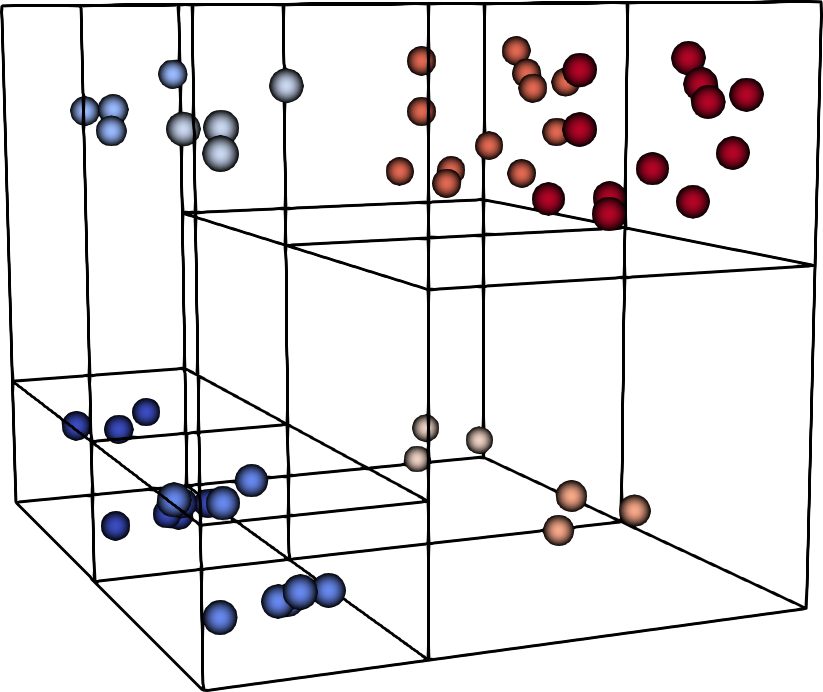}
    \caption{}
\end{subfigure}
\caption{(a) generic distribution of particles with clustering observed in CFD-DEM simulations and (b) partitioning
of particle clusters using orthogonal recursive bisection (ORB).}
\label{orb-fig}
\end{figure}

A generic distribution of particles that occur in a DEM simulation in a cartesian domain with clustering is shown in 
Fig.\ \ref{orb-fig}(a). A decomposition of these 
particles using ORB is shown in Fig.\ \ref{orb-fig}(b) where each subset contains the particle clusters. The explicit 
ORB scheme advances each of these domains separately with their local time scale to the desired global time. Therefore, 
domains with clustering where there are colliding particles perform multiple iterations while domains with particles that
may not collide advance with larger time-steps. There are three potential advantages to this approach;
\begin{enumerate}
\item The use of local time-step for colliding versus non-colliding particles reduces computational cost.
\item The neighbor search is restricted to each of the sub-domains as opposed to a global $O(n^2)$ search.
\item Advancing subsets of particles can improve temporal locality in memory  
as opposed to global update of all particles one after the other.
\end{enumerate}

\subsection{Numerical implementation}
In order to achieve spatial locality of fluid and particle data, the computational 
domain is divided into boxes that contain the fluid and co-located particles. Message passing 
interface (MPI) is used for parallelization where the boxes are distributed among 
various MPI ranks. The distribution of boxes are obtained in way that minimizes communication distance 
(a space-filling curve \cite{peano1890} approach is used) and equalizes load (a ``knapsack" \cite{rendleman2000} algorithm is 
applied to balance load). Each rank operate over the boxes they own and update fluid and particle fields with 
subsequent redistribution of ghost data. ORB strategy is used here to partition the
domain into boxes that balances the number of particles. 
The steps in the time advancement for the traditional explicit (with global minimum particle time-step) and the explicit ORB scheme 
is as shown in algorithms \ref{tradexpl-algo} and \ref{explorb-algo}, respectively. 
The detailed steps for the fluid update where the incompressible Navier-Stokes equations are solved, 
 have been skipped for brevity. Readers are referred to detailed algorithms and numerical methods 
presented by Syamlal et al. \cite{syamlal1993}.
\begin{algorithm}
 \caption{Traditional explicit scheme}
 \begin{algorithmic}
 \STATE Integrate fluid equations for time $\Delta t_f$\;
 \STATE Update particle drag forces\;
 \STATE $\Delta t_p$ = global minimum particle time-step
 \STATE $t_p = 0.0$
 \WHILE{$t_p < \Delta t_f$}
  \STATE nboxes = number of processor owned boxes
  \STATE update particle neighbor list\;
 \FOR{box $i$ in nboxes}
  \STATE find collision forces for particles within box $i$\;
  \STATE use velocity-verlet time-step update for time $\Delta t_p$\;
  \ENDFOR
  \STATE $t_p=t_p+\Delta t_p$\;
  \STATE redistribute ghost particles\;
  \ENDWHILE
  \STATE Deposit particle data on grid\;
  \STATE Regrid the computational domain using ORB\;
  \end{algorithmic}
  \label{tradexpl-algo}
\end{algorithm}
\begin{algorithm}
 \caption{explicit ORB scheme}
 \begin{algorithmic}
 \STATE nsubit = number of sub-iterations
 \STATE Integrate fluid equations for time $\Delta t_f$\;
 \STATE Update particle drag forces\;
 \STATE nboxes = number of processor owned boxes\;
 \STATE $\mathrm{\Delta t_{sp} = \frac{\Delta t_f}{nsubit}}$\;
 \STATE initialize array ``ltstep" of size nboxes\;
 \FOR{it in nsubit}
  \STATE update particle neighbor list\;
 \FOR{box $i$ in nboxes}
    \STATE ltstep[i]=minimum particle time-step within box $i$
 \ENDFOR
 \FOR{box $i$ in nboxes}
    \STATE $t_p = 0.0$
    \WHILE{$t_p < \Delta t_{sp}$}
        \STATE find collision forces for particles within box $i$\;
        \STATE use velocity-verlet time-step update for time, ltstep[i]\;
        \STATE $t_p=t_p+\mathrm{ltstep[i]}$\;
    \ENDWHILE
  \ENDFOR
  \STATE redistribute ghost particles\;
  \ENDFOR
  \STATE Deposit particle data on grid\;
  \STATE Regrid the computational domain using ORB\;
  \end{algorithmic}
  \label{explorb-algo}
\end{algorithm}

The traditional explicit method performs sub-iterations for the particle update over boxes 
with a global minimum time-step, while a local time-step for each box is used in the explicit ORB scheme. 
It should be noted that a global subcycling is also performed in the explicit ORB scheme, based on a user defined 
number of sub-iteration count (nsubit in Algorithm \ref{explorb-algo}). This is done so as to fine-grain the 
update of ghost particles and their redistribution among boxes thereby reducing errors due to lag in communication. 
The number of sub-iterations are on the order of 5-40 for stable 
particle time integration; its sensitivity to errors in solution is studied in section \ref{results-sec} for different 
particle distribution scenarios.
The local time-step for a particle is assigned as the collisional time scale scaled by Courant number of 0.02, when there are 
potential collision partners within a distance of 3 particle radii ($dt=\tau_p/50$) . The sub time-step, $\Delta t_{sp}$ (see Algorithm \ref{explorb-algo}), 
obtained from the user defined sub-iteration count, is used as local time-step for particles that have no collision partners. 
The boxes are redefined (regrid) using ORB at the end of the fluid and particle update so as to capture new particle clusters
in the subsequent time-step. 

The use of local time-step for the boxes owned by processors introduces load imbalance in terms of number of time-steps
performed per processor; boxes with particle clusters perform more number of time-steps compared to dilute regions. Therefore,
the ``knapsack" algorithm is used to rebalance loads among processors after ORB partitioning, based on 
local time-stepping costs associated with each box.

%% file: results.tex
\section{Results and discussion}
\label{results-sec}
The computational performance of the explicit ORB scheme is compared with the traditional
explicit method in this section for various test cases in both serial and parallel execution of the algorithm. 
All the cases studied in this section were run on Intel Haswell 2.3 GHz processors \cite{haswell} with 24 cores per node. 
The sensitivity of solution to the number of sub-iterations (nsubit in Algorithm \ref{explorb-algo}) 
in the particle update for the explicit ORB scheme is studied in detail 
and an optimal number for different scenarios is inferred.
\section{Serial cases}
\subsection{Homogenous cooling system (HCS)}
This system consists of a cartesian domain with periodic boundaries containing particles initialized
with a gaussian distribution of velocity and random position vectors. 
The total energy of the particles in this system decreases 
over time due to collisional and gas-solid drag losses. 
This idealized case has an analytic solution for average energy decay  
and such a system tends to arise in regions of gas-solid flows where there is an isotropic 
distribution of particles.
\begin{figure}
\centering
\begin{subfigure}{0.48\textwidth}
\includegraphics[width=0.95\textwidth]{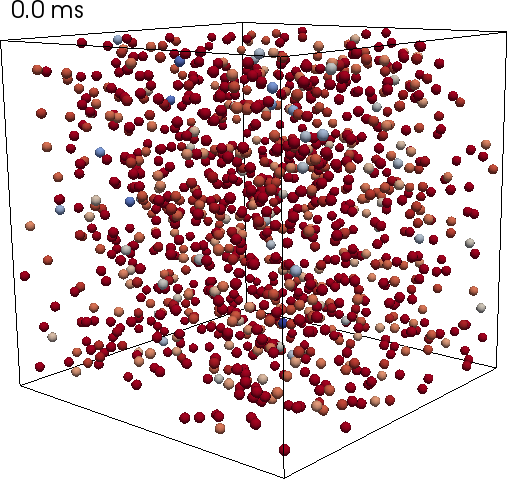}
\caption{}
\end{subfigure}
\begin{subfigure}{0.48\textwidth}
\includegraphics[width=0.95\textwidth]{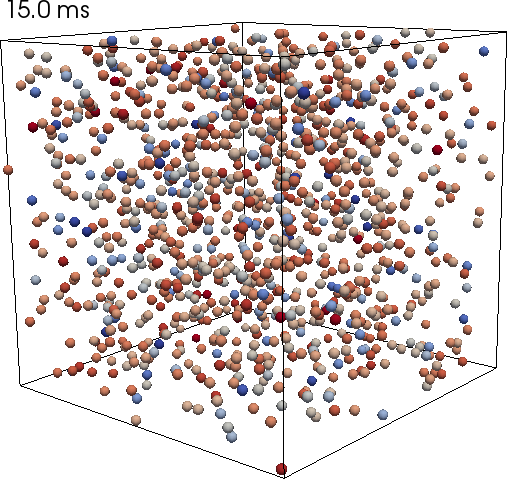}
\caption{}
\end{subfigure}
\begin{subfigure}{0.48\textwidth}
\includegraphics[width=0.95\textwidth]{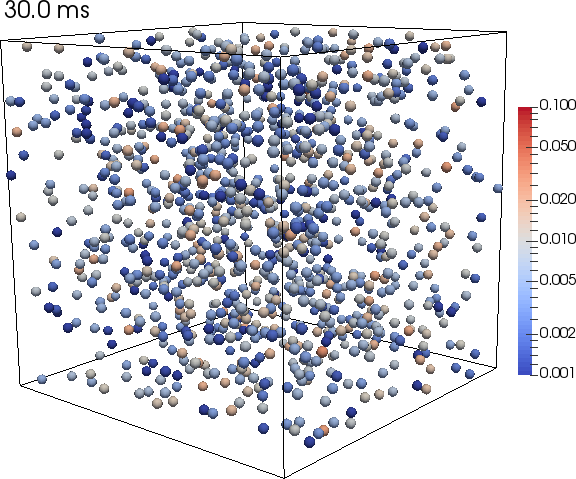}
\caption{}
\end{subfigure}
\begin{subfigure}{0.48\textwidth}
\includegraphics[width=0.95\textwidth]{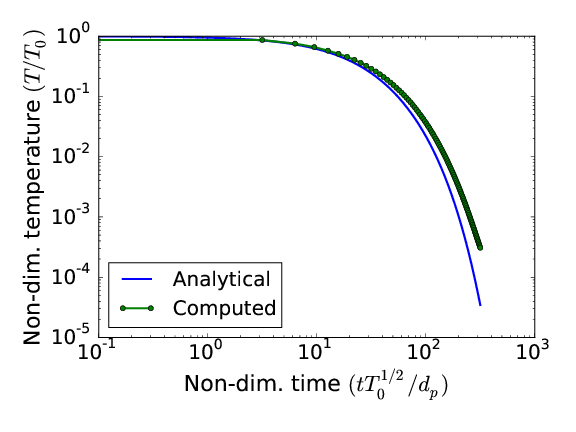}
\caption{}
\end{subfigure}
\caption{Snapshots of particle specific kinetic energy ($k=\frac{1}{2} |\mathbf{v}|^2$) in $\mathrm{m^2/s^2}$  
for the HCS simulation at (a) 0 , 
(b) 15 and (c) 30 ms, respectively. (d) shows the simulated and analytic solutions for particle averaged 
temperature ($T=\frac{1}{3}\overline{\mathrm{v}^2}$) decay as a function of time. Here, $T$ is normalized by the initial 
temperature, $T_0$ and time is normalized by $\frac{d_p}{\sqrt{T_0}}$ where $d_p$ is the particle diameter.}
\label{hcs-fig}
\end{figure}
\begin{table}
\centering
\begin{tabular}{|c|c|}
\hline
Simulation parmeters & Value \\
\hline
Particle diameter ($d_p$)   & $\mathrm{100}~\mu m$ \\
Particle density ($\rho_p$) & $\mathrm{1000~kg/m^3}$ \\
Gas density ($\rho_g$)      & $\mathrm{1~kg/m^3}$\\
Gas viscosity ($\mu_g$)     & $\mathrm{2\times 10^{-5}~Pa~sec}$ \\
Collisional spring constant ($k$) & $\mathrm{10~N/m}$ \\
Restitution coefficient ($e$) & 0.8 \\
Fluid time-step ($\Delta t_f$) & 0.0001 sec \\
Drag model & BVK model \cite{beetstra2007}\\
\hline
\end{tabular}
\caption{Simulation parameters for the homogenous cooling system case}
\label{hcs-tab}
\end{table}
Fig.\ \ref{hcs-fig}(a), (b) and (c) show snapshots of particles colored by their kinetic 
energy at 0, 15 and 30 ms for a typical HCS simulation performed in a cubical domain of size 4~mm with 1222 particles, 
indicating the decay of energy over time. The mesh consists of 8000 cells with 20 cells along each coordinate direction. 
The material parameters for the gas and solid phase are described in Table \ref{hcs-tab}. 

Fig.\ \ref{hcs-fig}(d) shows the comparison 
between the computed and analytic solutions for the temperature ($T=\frac{1}{3}\overline{\mathrm{v}^2}$) 
decay as predicted by the Haff's law \cite{haff1983,fullmer2017}, indicating reasonable agreement. Slight deviations are 
observed due to the idealized assumptions (such as constant collision cross-section) used in the analytic derivation of 
Haff's law.
\begin{figure}
\centering
\begin{subfigure}{0.48\textwidth}
\includegraphics[width=0.95\textwidth]{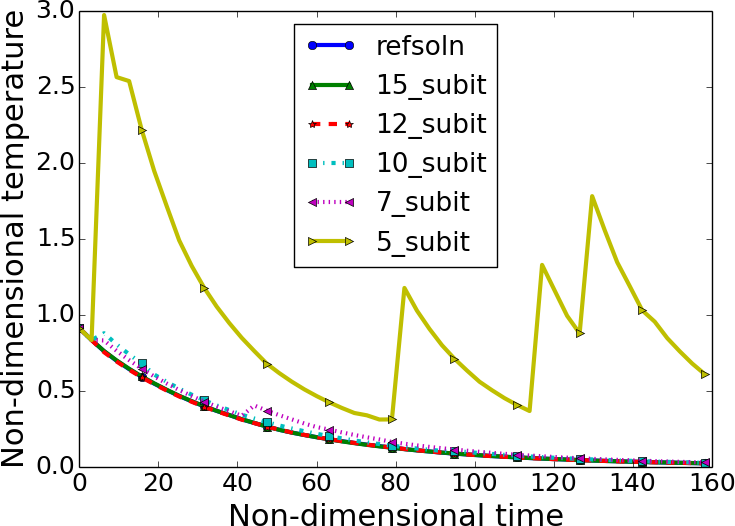}
\caption{}
\end{subfigure}
\begin{subfigure}{0.48\textwidth}
{\includegraphics[width=0.95\textwidth]{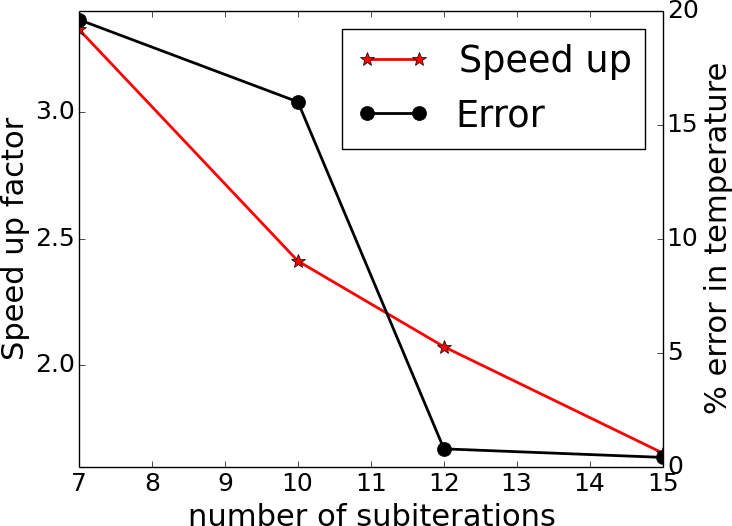}}
\caption{}
\end{subfigure}
\begin{subfigure}{0.48\textwidth}
{\includegraphics[width=0.95\textwidth]{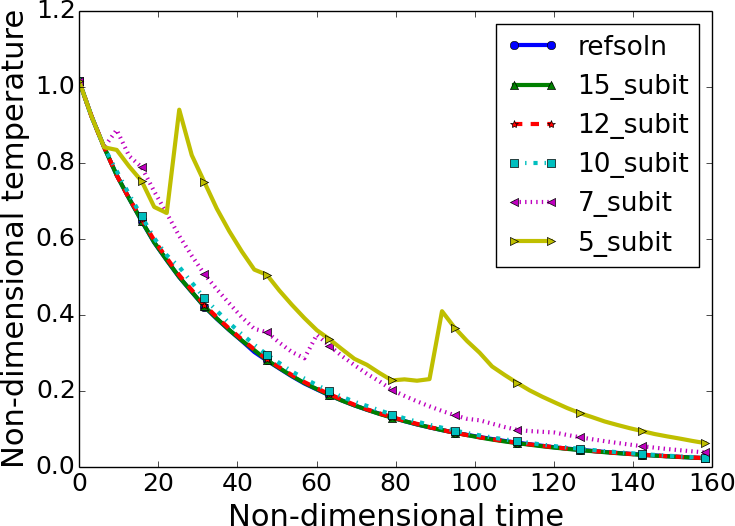}}
\caption{}
\end{subfigure}
\begin{subfigure}{0.48\textwidth}
{\includegraphics[width=0.95\textwidth]{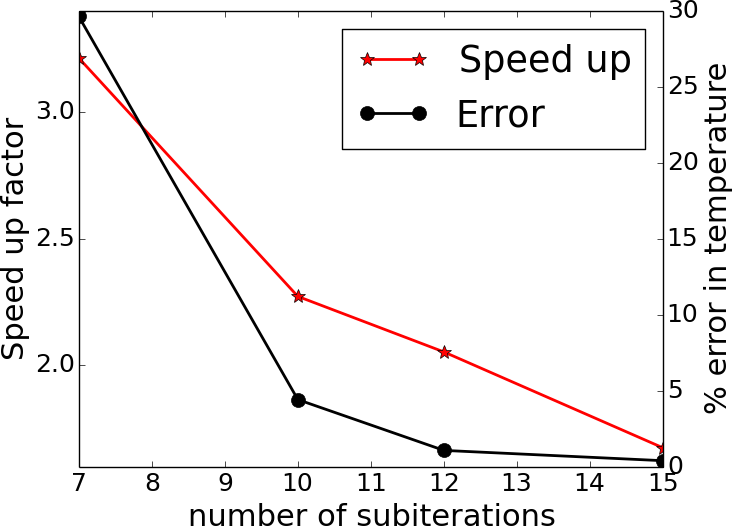}}
\caption{}
\end{subfigure}
\begin{subfigure}{0.48\textwidth}
{\includegraphics[width=0.95\textwidth]{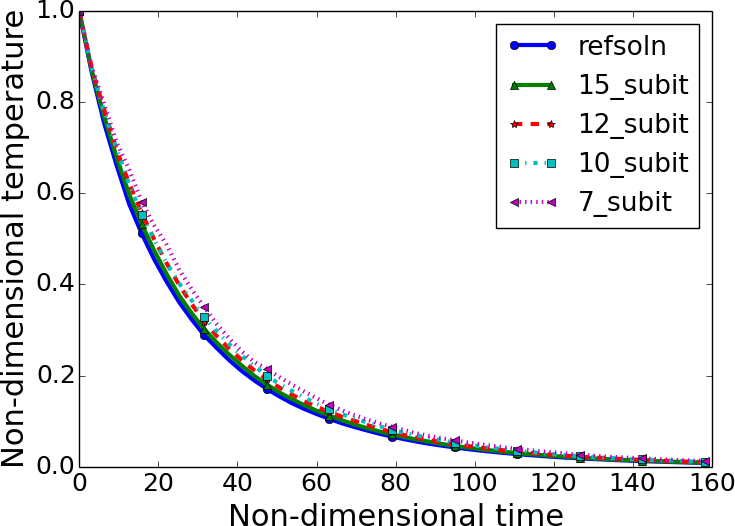}}
\caption{}
\end{subfigure}
\begin{subfigure}{0.48\textwidth}
{\includegraphics[width=0.95\textwidth]{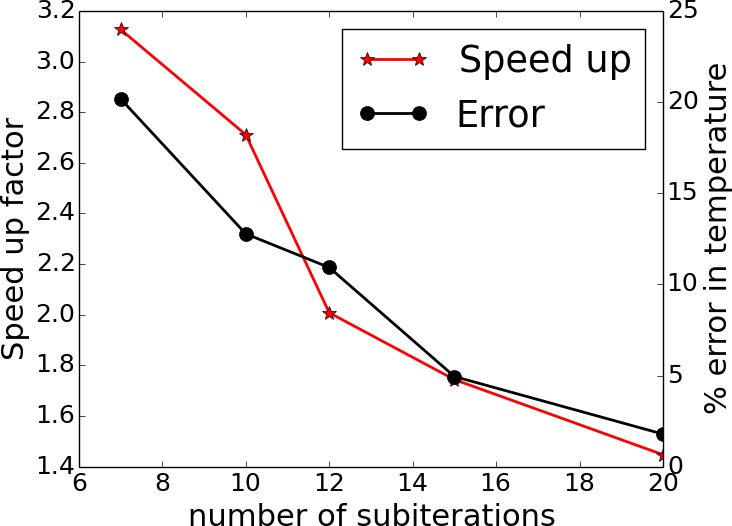}}
\caption{}
\end{subfigure}
\caption{Computed solution of temperature decay using the explicit ORB method with varying number of sub-iterations for 
(a) 100, (c) 200 and (e) 1222 particles, respectively. Figures (b), (d) and (f) show 
the variation of error in non-dimensional temperature between the explicit ORB and the traditional
explicit method along with the speed-up achieved for varying number of sub-iteration counts.}
\label{hcserr-fig}
\end{figure}

Fig.\ \ref{hcserr-fig} compares the non-dimensional temperature decay predicted by the traditional explicit and the explicit ORB scheme for 
varying number of particle and sub-iteration counts.
The traditional explicit scheme solution is assumed to be the reference and 
an average temperature error is obtained for the explicit ORB method for varying number of 
sub-iteration counts. The error is further normalized by the time-averaged temperature so as to quantify the fractional deviation 
in the solution. 

Fig.\ \ref{hcserr-fig}(a), (c) and (e) show temperature decay solutions for varying number of particles 
in the domain from 100 to 1222 particles. 4 ORB levels (16 leaves) are used 
for the lower particle count cases (100 and 200) while 6 levels (64 leaves) are used for the simulation with 
1222 particles. Large deviations from the reference solution is observed for 
lower number of sub-iterations in the explicit ORB scheme. 
The larger the sub time-step, the greater is the chance of missing collisions 
thereby increasing the error. Fig.\ \ref{hcserr-fig}(b), (d) and (f) show the variation of 
the speed-up factor and the error as a function of number of sub-iterations. 
As the number of sub-iterations is increased, the 
explicit ORB method tends towards the traditional explicit scheme and its performance is reduced while 
lower number of sub-iterations result in greater error in the solution. 
Nonetheless, a nominal speed-up factor of 2X is obtained with 10-15 sub-iterations for all the 3 cases, 
with errors in the range of 1-5\% with respect to the traditional explicit method.
\subsection{Particle settling problem}
The settling of particles due to gravity in a cartesian domain is studied in this section. 
The particle distribution here transitions from being uniformly distributed to a settled bed that is
densely packed, thereby testing the performance and accuracy of the explicit ORB scheme 
with time varying particle densities. 
\begin{figure}
\centering
\begin{subfigure}{0.42\textwidth}
\includegraphics[width=0.99\textwidth]{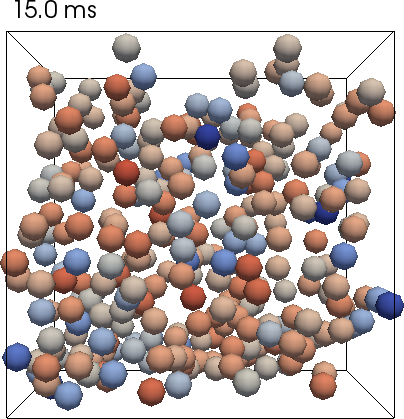}
\caption{}
\end{subfigure}
\begin{subfigure}{0.42\textwidth}
\includegraphics[width=0.99\textwidth]{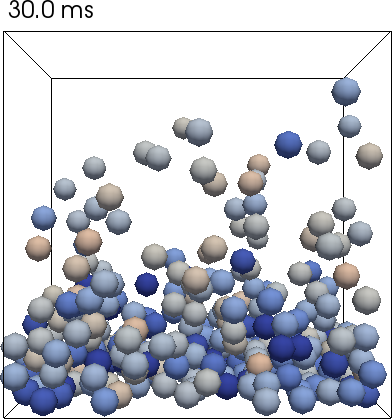}
\caption{}
\end{subfigure}
\begin{subfigure}{0.42\textwidth}
\includegraphics[width=0.99\textwidth]{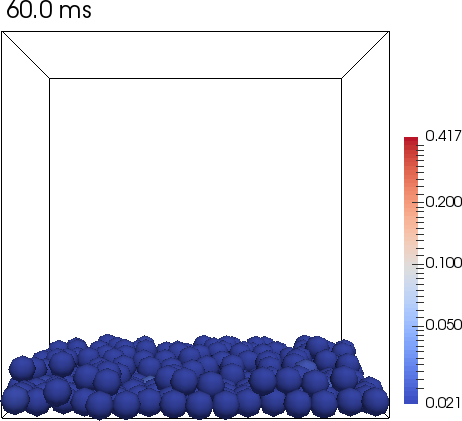}
\caption{}
\end{subfigure}
\begin{subfigure}{0.48\textwidth}
\includegraphics[width=0.99\textwidth]{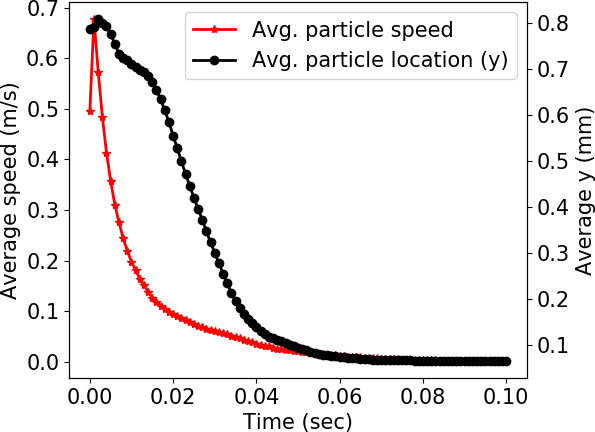}
\caption{}
\end{subfigure}
\caption{Snapshots of particle speed ($|\mathbf{v}|$) in $\mathrm{m/s}$  
for the particle settling simulation at (a) 15 , 
(b) 30 and (c) 60 ms, respectively. (d) shows the particle averaged position along the direction of gravity (y) 
and speed vs time.}
\label{set-fig}
\end{figure}
Fig.\ \ref{set-fig} illustrates the physics of the settling simulation through time snapshots and particle 
averaged parameters. This simulation consists of 300 particles in a cubical box, 1.5~mm in size. 
The mesh consists of 1000 cells with 10 cells along 
each coordinate direction. The material properties used in this simulation is the same as 
the HCS case, as described in Table \ref{hcs-tab}.
Gravity is assumed to be along the y coordinate direction (top to bottom).   
The particles that are uniformly distributed initially, settle and lose their gravitational 
potential energy through drag, inter-particle and wall collisions. 
Fig.\ \ref{set-fig}(a) to (c) show particle distribution colored by their speeds from 15 to 60 ms; 
the particle kinetic energy is observed to dissipate completely at the end of 60 ms. 
Fig.\ \ref{set-fig}(d) shows the average particle speed and 
position along the y direction; the particle speeds are observed to rise initially during free fall and assymptotes to 0 
after longer time periods.
\begin{figure}
\centering
\begin{subfigure}{0.48\textwidth}
\includegraphics[width=0.95\textwidth]{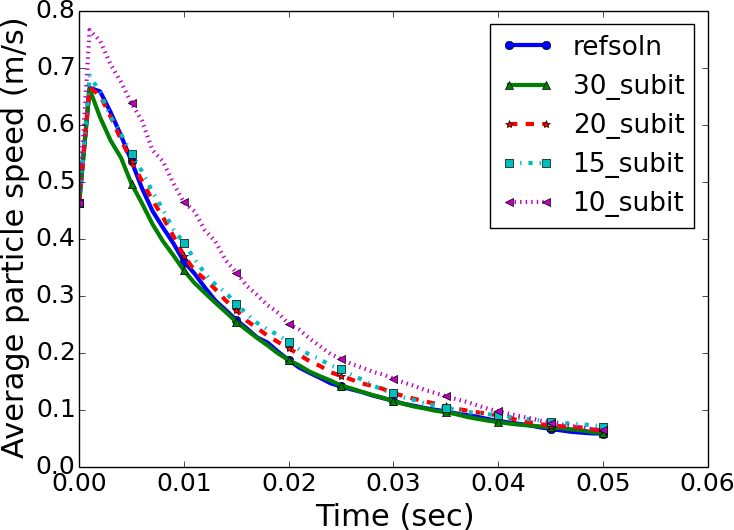}
\caption{}
\end{subfigure}
\begin{subfigure}{0.48\textwidth}
\includegraphics[width=0.95\textwidth]{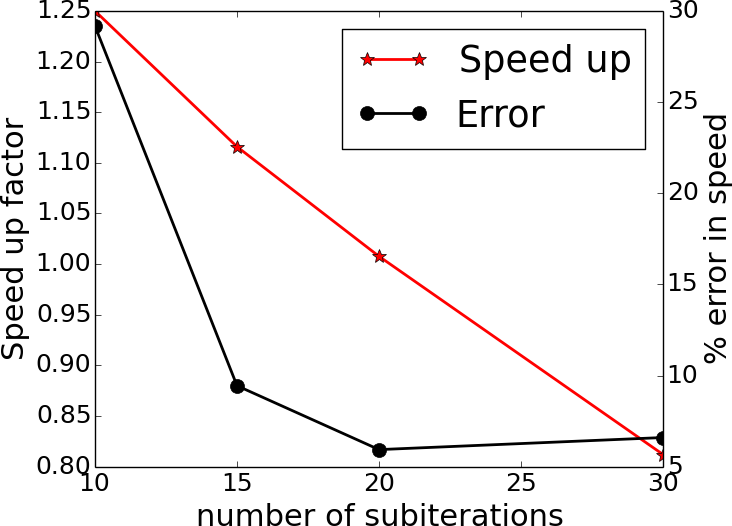}
\caption{}
\end{subfigure}
\begin{subfigure}{0.48\textwidth}
\includegraphics[width=0.95\textwidth]{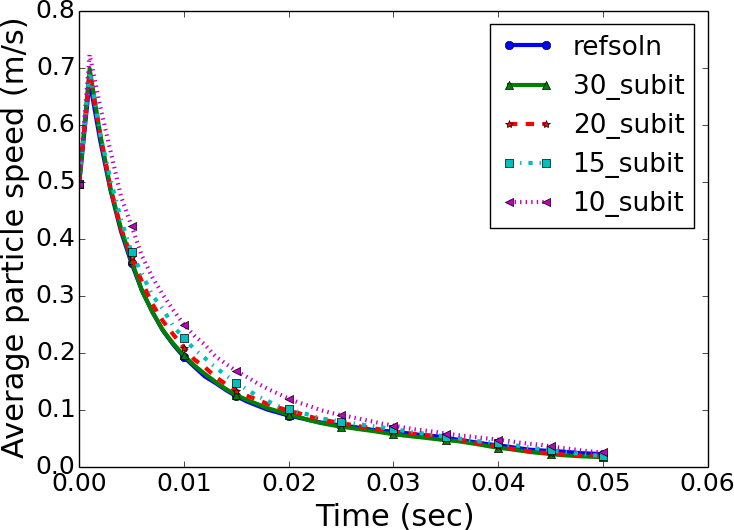}
\caption{}
\end{subfigure}
\begin{subfigure}{0.48\textwidth}
\includegraphics[width=0.95\textwidth]{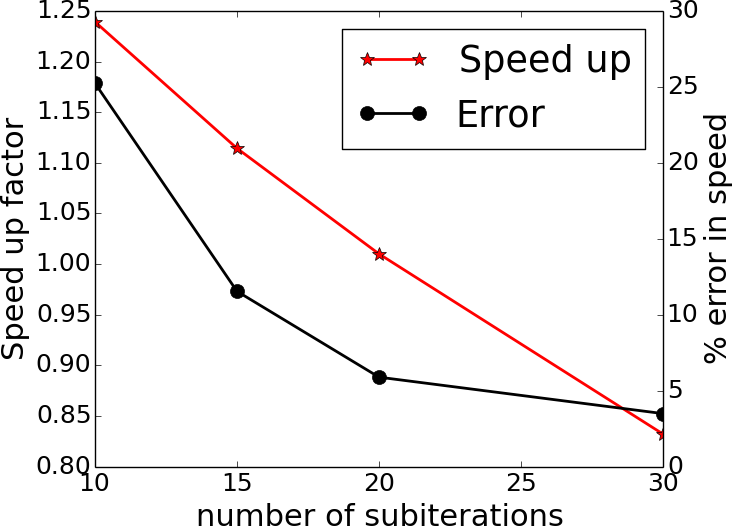}
\caption{}
\end{subfigure}
\caption{Computed solution of average particle speed using the traditional explicit and explicit ORB method 
with varying number of sub-iterations for 
(a) 100 and (c) 300, respectively. (b) and (d) show the variation of error in particle averaged speed 
between the explicit ORB and the traditional explicit method 
along with the speed-up achieved for varying number of sub-iteration counts.}
\label{seterr-fig}
\end{figure}

Fig.\ \ref{seterr-fig} shows the effect of sub-iteration count in the explicit ORB scheme 
on the solution accuracy. Fig.\ \ref{seterr-fig}(a) and (c) show the 
average particle speed as a function of time for simulations with 100 and 300 particles, 
respectively, for varying number of sub-iteration counts. These simulations are performed with 4 ORB levels (16 leaves). 

A similar trend with respect to the 
HCS test case is observed; larger number of sub-iterations reduce the error in solution as shown 
in Fig.\ \ref{seterr-fig}(b) and (d), respectively. The error in the explicit ORB 
solution (Fig.\ \ref{seterr-fig}(b) and (d)) is computed based on the average particle speed 
variation with respect to the reference traditional explicit scheme 
solution. The speed-up factor achieved with the 
use of explicit ORB scheme is lower in this case due to the highly collisional distribution 
of particles. There is no speed-up observed using the explicit ORB scheme for maintaining solution 
errors within 1-5\%. 

The explicit ORB scheme thus performs better when there are dilute regions in the domain. 
The settling of particles in this problem rapidly transitions the system into a dense distribution,
thereby enforcing small particle time-steps in all boxes. The traditional explicit scheme is 
recovered for dense distributions where a global small time-step is the same as the local time-step.
\section{Parallel cases}
\subsection{Riser flow}
\label{riser-sec} 
This test case is representative of industrial systems such as circulating 
fluidized bed (CFB) riser reactors \cite{zhang2001} that are applied 
to catalytic cracking and combustion.
This simulation consists of a rectangular column with a 6~mm x 6~mm cross section and a height of 1~cm, with 
14,000 particles seeded uniformly in the domain with periodic boundary conditions at all faces. 
A constant pressure gradient of 2~Pa along the axial direction is imposed.  
Simulations are performed on 64 MPI ranks with 7 ORB levels (maximum of 128 leaves) on a mesh of 
51,200 cells (32 x 50 x 32 grid). The material properties from Table \ref{hcs-tab} along with 
the Koch-Hill drag model \cite{hill2001} is used in these simulations.
\begin{figure}
\centering
\begin{subfigure}{0.255\textwidth}
\includegraphics[width=0.95\textwidth]{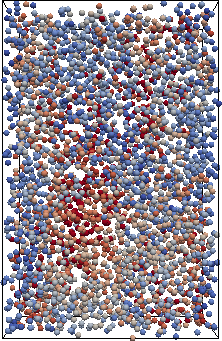}
\caption{}
\end{subfigure}
\begin{subfigure}{0.32\textwidth}
\includegraphics[width=0.95\textwidth]{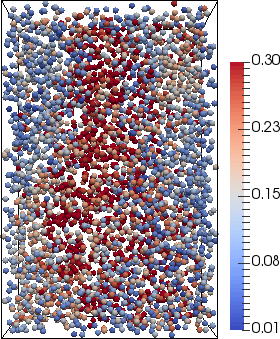}
\caption{}
\end{subfigure}
\begin{subfigure}{0.3\textwidth}
\includegraphics[width=0.95\textwidth]{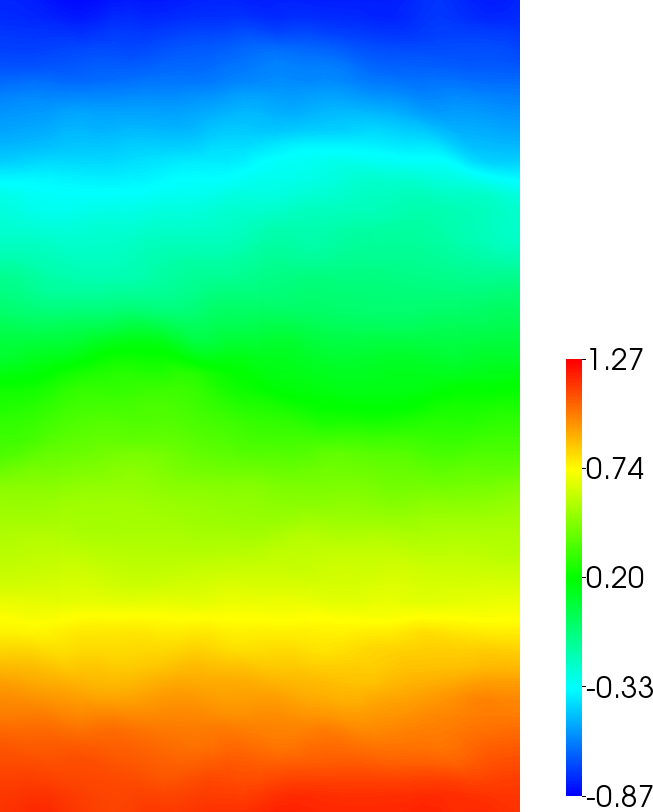}
\caption{}
\end{subfigure}
\caption{Snapshots of particle distribution colored by their speed in m/s for the riser flow simulation 
at (a) 300 ms and (b) 400 ms, respectively. 
(c) shows the distribution of pressure (Pa) in the domain at t=400 ms.}
\label{risersnapshots-fig}
\end{figure}
Fig.\ \ref{risersnapshots-fig} shows the snapshots of particle distribution and 
fluid pressure; the favorable bottom-to-top pressure gradient facilitates movement of 
particles predominantly along the axis. The particles are recycled back into the domain via 
the periodic boundaries and are steadily accelerated due to the constant pressure gradient.
\begin{figure}
\centering
\begin{subfigure}{0.45\textwidth}
\includegraphics[width=0.99\textwidth]{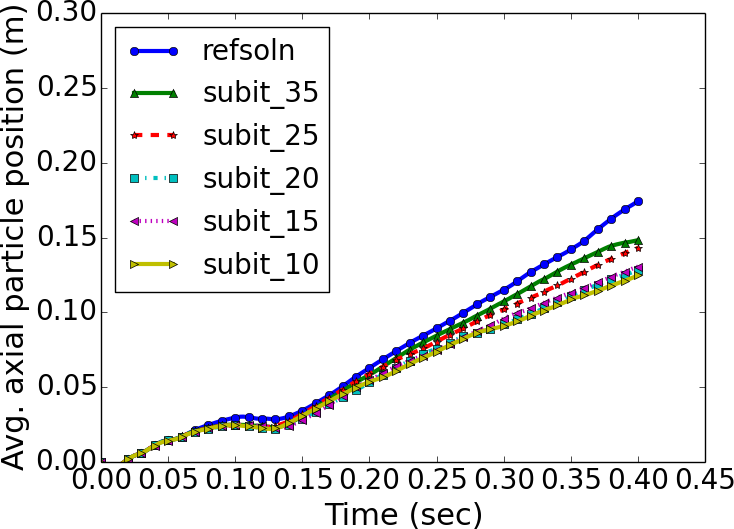}
\caption{}
\end{subfigure}
\begin{subfigure}{0.45\textwidth}
\includegraphics[width=0.99\textwidth]{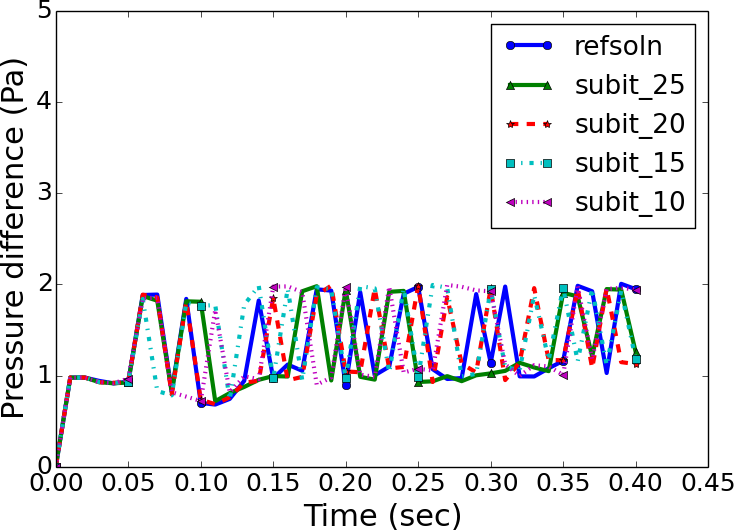}
\caption{}
\end{subfigure}
\begin{subfigure}{0.45\textwidth}
\includegraphics[width=0.99\textwidth]{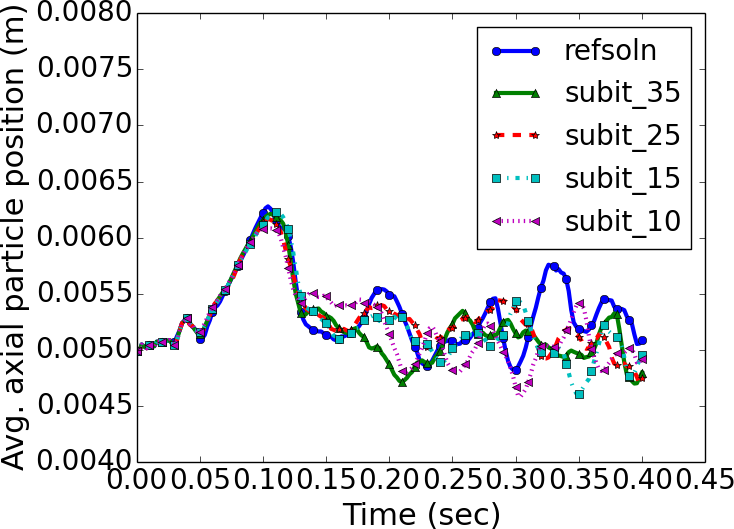}
\caption{}
\end{subfigure}
\begin{subfigure}{0.45\textwidth}
\includegraphics[width=0.99\textwidth]{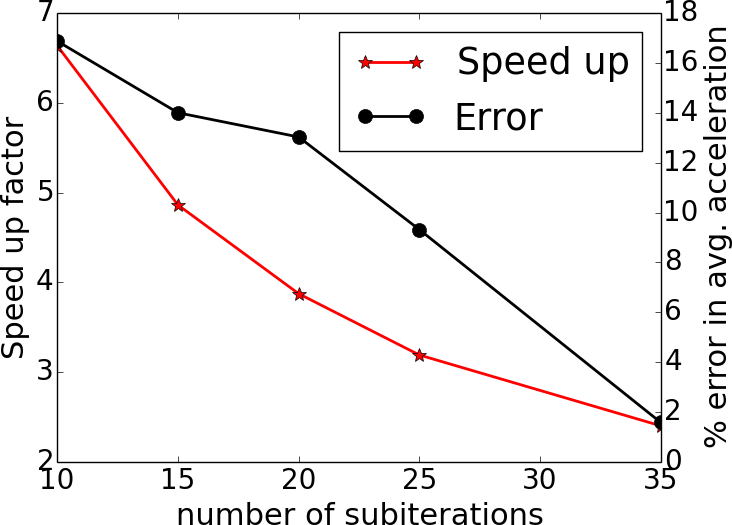}
\caption{}
\end{subfigure}
\caption{(a) particle averaged speed along the axial direction, (b) pressure difference between riser top and bottom 
and (c) is the particle averaged position along the axis as a function of time and varying sub-iteration counts.
(d) plots the variation of speed-up and error in particle averaged acceleration with respect to the traditional explicit method 
for varying number of sub-iterations in the explicit ORB scheme.}
\label{riser_err-fig}
\end{figure}

The average particle speed along the axial direction as a function of time is shown in Fig.\ \ref{riser_err-fig}(a) for the reference traditional 
explicit time-stepping as well as the explicit ORB scheme with varying number of sub-iterations. 
The average speeds are observed to approach a constant slope with respect to time indicating a steady-state 
acceleration. There is reasonable agreement among the explicit ORB cases  
with higher error for lower number of sub-iteration counts. Fig.\ \ref{riser_err-fig}(b) shows
the pressure difference between the top and bottom of the column as a function of time for this case. 
This parameter approaches a fluctuating steady-state within 0.15 seconds; the explicit ORB scheme 
is able to achieve the averaged steady-state value for 
all sub-iteration counts. Fig.\ \ref{riser_err-fig}(c) shows the average particle position 
along the axial direction as a function of time. A fluctuating steady-state about the 
middle of the column (0.005 m from the bottom) is attained for all the cases with reasonable agreement.
Fig.\ \ref{riser_err-fig}(d) shows the sensitivity of sub-iteration count on solution accuracy and the speed-up
obtained using the explicit ORB method. The steady-state acceleration is chosen as the parameter of interest for computing errors
with respect to the reference traditional explicit method. The errors are within 1-2\% for about 35 sub-iterations along with  
a 2X speed-up using the explicit ORB scheme. 
\subsection{Fluidized bed}
\label{flubed-sec}
This case consists of a rectangular column with a 1.6~mm x 1.6~mm cross section and a height of 1~cm, 
with 10,000 particles seeded initially close to the bottom of the column, on a mesh of 51,200 cells (32 x 50 x 32 grid). 
A constant velocity inflow boundary condition is applied to the bottom face with a fixed normal gas velocity of 1.5~cm/s; 
the lateral surfaces are assumed to be no-slip walls and pressure outflow boundary condition is used at the top. 
This simulation uses the same material properties as in the riser case (section \ref{riser-sec}) and 
is perfomed using 16 MPI ranks with 6 ORB levels (64 leaves). 
\begin{figure}
\centering
\begin{subfigure}{0.096\textwidth}
\includegraphics[width=0.95\textwidth]{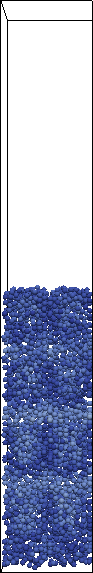}
\caption{}
\end{subfigure}
\begin{subfigure}{0.1\textwidth}
\includegraphics[width=0.95\textwidth]{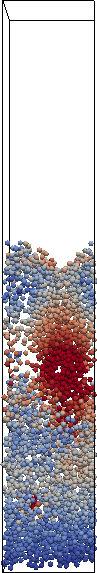}
\caption{}
\end{subfigure}
\begin{subfigure}{0.1\textwidth}
\includegraphics[width=0.95\textwidth]{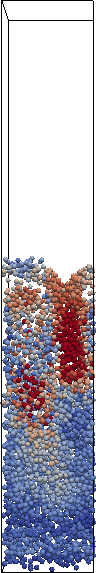}
\caption{}
\end{subfigure}
\begin{subfigure}{0.18\textwidth}
\includegraphics[width=0.95\textwidth]{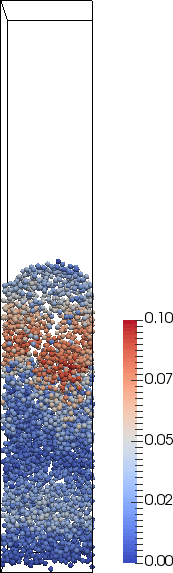}
\caption{}
\end{subfigure}
\caption{Snapshots of particle distribution colored by their speed 
in m/s for the fluidized bed simulation 
at (a) 0, (b) 100, (c) 200 and (d) 400 ms, respectively.}
\label{flubedsnapshots-fig}
\end{figure}

Fig.\ \ref{flubedsnapshots-fig} shows the snapshots of 
particle distribution in the domain at four time points between 0 and 400 ms, indicating 
top-to-bottom mixing, typically seen in fluidized bed reactors.
\begin{figure}
\centering
\begin{subfigure}{0.45\textwidth}
\includegraphics[width=0.99\textwidth]{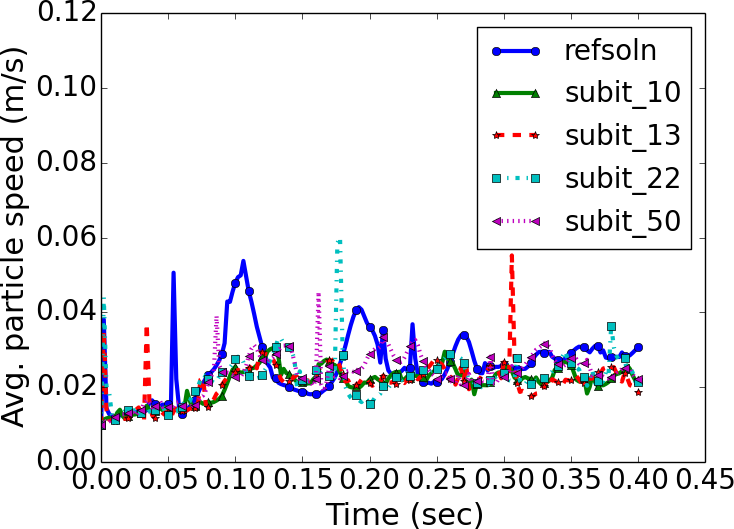}
\caption{}
\end{subfigure}
\begin{subfigure}{0.45\textwidth}
\includegraphics[width=0.99\textwidth]{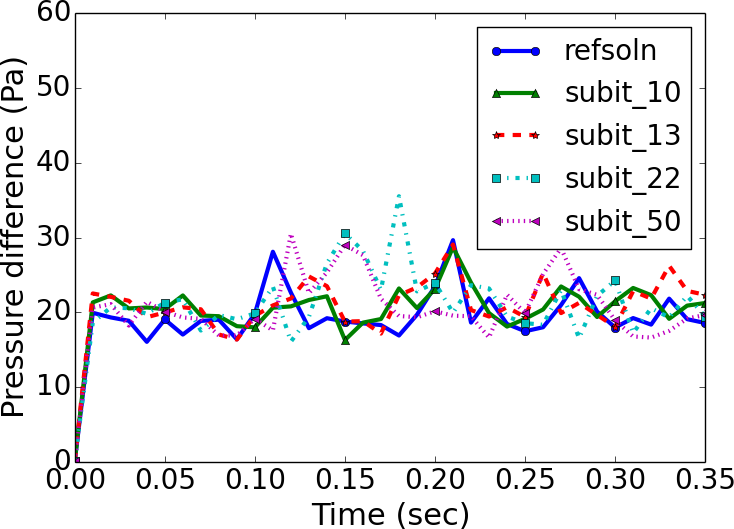}
\caption{}
\end{subfigure}
\begin{subfigure}{0.45\textwidth}
\includegraphics[width=0.95\textwidth]{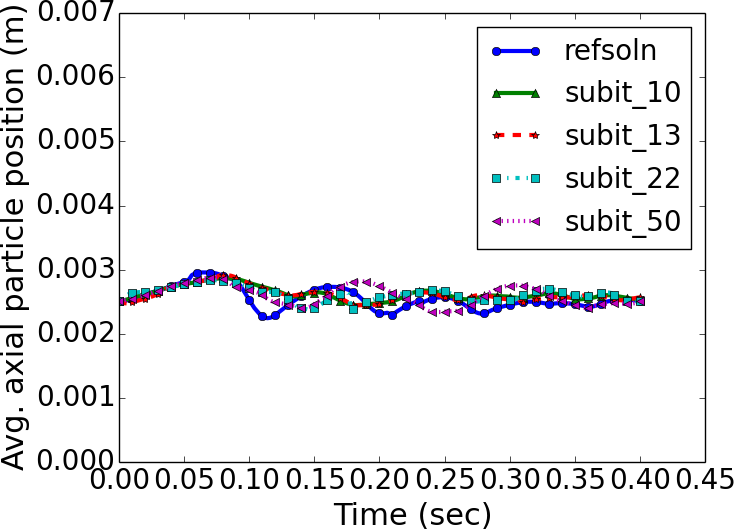}
\caption{}
\end{subfigure}
\begin{subfigure}{0.45\textwidth}
\includegraphics[width=0.95\textwidth]{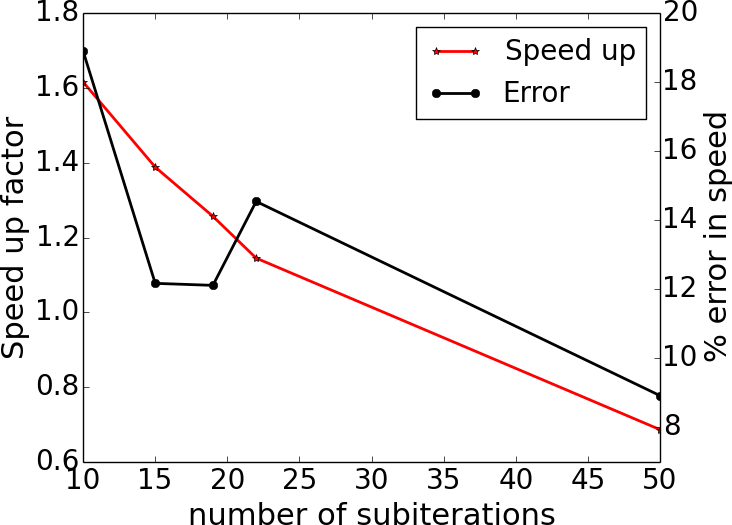}
\caption{}
\end{subfigure}
\caption{(a) particle averaged speed, (b) top to bottom pressure difference
and (c) particle averaged axial position in the fluidized bed as a 
function of time for the traditional explicit and the  explicit ORB scheme 
with varying sub-iteration counts. (d) is the variation of speed-up and 
error in steady-state particle averaged speed solution for varying number of sub-iterations.}
\label{flubed_err-fig}
\end{figure}

Fig.\ \ref{flubed_err-fig} quantifies the accuracy and performance of the 
explicit ORB scheme for varying number of sub-iteration counts.
The fluidized bed is a chaotic system with dense particle distributions; a fluctuating steady-state for the particle speeds
is achieved in about 100 ms (Fig.\ \ref{flubed_err-fig}(a)) with intermittent peaks in average particle speeds 
over time. The explicit ORB scheme solutions show comparable trends with respect 
to the traditional explicit scheme and  approach similar averaged particle speeds at steady-state.
Fig.\ \ref{flubed_err-fig}(b) shows the variation of pressure difference between the top and bottom of the column; a fluctuating 
steady-state is attained within 50 ms. All of the explicit ORB scheme cases exhibit the same trend as the reference explicit solution 
with the nominally similar average pressure drop at steady-state.
The variation of average axial particle position with time is shown Fig.\ \ref{flubed_err-fig}(c) indicating agreement among 
the explicit ORB scheme cases with respect to the reference solution.

Fig.\ \ref{flubed_err-fig}(d) shows the variation of speed-up and error in 
steady-state time-averaged particle speed for varying sub-iteration counts in the explicit ORB scheme. 
A speed-up of 1.1-1.3X is seen for 12-20 sub-iteration counts with solution errors on the order of 12-14\%. 
The number of sub-iterations needs to be around 50 to bound the errors within 10\% for which no speed-up is observed. 

This case reinforces the need for dilute regions in the problem for optimal 
performance of the explicit ORB scheme. The dense particle distributions in this problem will require the global resolution 
of collisional time scales among all particles thereby negating any performance improvements from the explicit ORB scheme.

%% file: conclusions.tex
\section{Conclusions and future work}
An adaptive timestepping method is developed to speed up CFD-DEM simulations 
using an orthogonal recursive bisection based approach. This algorithm was implemented
in a parallel CFD-DEM solver and its performance was compared against the baseline explicit
scheme with a global time step.
Four different test cases with dilute and dense particle distributions were studied 
to quantify the efficacy of this method. The dilute cases (homogenous cooling system and riser flow) 
showed a 2-3X speed-up relative to baseline explicit scheme, while maintaining minimal
solution errors. The dense cases on the other hand (settling and fluidized bed) showed 
minimal speed-up to maintain low solution errors. The current method provides a means of 
setting the timestep appropriately on a particle cluster basis, but caution 
should be used away from collisions to ensure that accuracy requirements 
(e.g., based on path line integration considerations) are still met.

Other bisection strategies such as k-means clustering that respect clustering and probable 
collision partners will be studied in the context of relatively denser distributions. A relative 
distance based time step estimate instead of a bimodal distribution used in this work, will be considered
to prevent collision misses in dense systems.